\documentclass[11pt,a4paper]{JHEP3}
\usepackage{amssymb}
\usepackage{epsfig,amssymb,euscript}

\def\d{\partial}
\def\k{\kappa}
\def\1{{\bf 1}}

\def\d{\partial}

\def\a{\alpha}

\def\0{\nonumber}
\newcommand{\tU}{{\tilde U}}
\newcommand{\tC}{{\tilde C}}

\newcommand{\tN}{{\tilde N}}

\newcommand{\tV}{{\tilde V}}
\newcommand{\tE}{{\tilde E}}
\newcommand{\tT}{{\tilde T}}
\newcommand{\tS}{{\tilde S}}

\newcommand{\Q}{{\mathcal Q}}
\newcommand{\N}{{\mathcal N}}

\newcommand{\V}{{\mathcal V}}

\newcommand{\Y}{{\mathcal Y}}

\def\beq{\begin{equation}}
\def\eeq{\end{equation}}
\def\eea{\end{eqnarray}}     %eqnarray
\def\bea{\begin{eqnarray}}

\def\e{{\rm e}}

\font\doppio=msbm10 at11pt
\newcommand{\RR}{\hbox {\doppio R}}

\preprint{SISSA/55/03/EP  \\\tt hep-th/0306252}

\title{Star Democracy in Open String Field Theory}
\author {C. Maccaferri, D. Mamone \\
International School for Advanced Studies (SISSA/ISAS)\\
Via Beirut 2--4, 34014 Trieste, Italy, and INFN, Sezione di
Trieste\\
E-mail:   \email{maccafer@sissa.it}, \email{mamone@sissa.it}}

\abstract{We study three types of star products in SFT: the ghosts, the twisted ghosts and the  matter. We find that
 their Neumann coefficients are related to each other in a compact way which includes the Gross-Jevicki relation between
 matter and ghost sector: we explicitly show that the same relation, with a minus sign, holds for the twisted and
 nontwisted ghost (which are different but define the same solution). In agreement with this, we prove that matter
 and twisted ghost coefficients just differ  by a minus sign. As a consistency check, we also compute the spectrum
 of the twisted ghost vertices from conformal field theory and,  using equality of twisted and reduced slivers, we
 derive the spectrum of the non twisted ghost star.}

\keywords{String Field Theory, Ghosts}

\begin{document}

%%%%%%%%%%%%%%%%%%%%%%%%%%%%%%%%%%%%%%%%%%%%%%%
\section{Introduction}
%%%%%%%%%%%%%%%%%%%%%%%%%%%%%%%%%%%%%%%%%%%%%%%
Open Bosonic String Theory is made up with two sectors, the embedding coordinates in the 26
 dimensional target space (which we call the matter), and a $bc$ system with conformal weights $(2,-1)$ (which
 we call the ghost).\\
The second quantized version is known to be  Witten String Field Theory \cite{W1}. In this theory the interaction
between string fields is encoded in the star product which defines an associative non commutative  multiplication
in the algebra of string fields.\\
This theory has lead to a  successful study of the process of tachyon condensation \cite{Sen, ohmori, WT, WT0},
 and hence to the identification of a stable vacuum of open string theory.\\
The theory which is supposed to describe strings  around the stable vacuum were postulated in \cite{RSZ1}, under
the name of ``Vacuum String Field Theory''(VSFT).
Such theory is formally identical to Witten's one, except for the kinetic operator, which is no more the
usual BRST charge, but a $c$-midpoint insertion. In such a theory the ghost sector is completely decoupled
from the matter, hence solutions can be found in a factorized form $matter\otimes ghost$.
Such solutions were obtained following two
parallel methods: one which is algebraic and is based on the oscillator
expansion of the string field \cite{KP, HKw, RSZ1, RSZ2}, the other is
based on the Boundary Conformal Theory which describes the original
unstable D--brane configuration \cite{GRSZ1, GRSZ2} .\\
The two approaches, although very different in the formalism, were shown to
lead to the  same results, first numerically by level expansion analysis, and
then analytically in \cite{okuda, Oku2} by making  use of the continuous basis of the star product.\\
Even if  the ghost sector of the theory is decoupled from the matter, nevertheless it is crucial for the
  consistency of VSFT, since
 it is supposed to contain all the tree level dynamics of open strings at the tachyon vacuum. For this reason
  a careful study of this sector is needed in order to understand better what is the physics behind it. One of the aims
   of this paper is indeed to  shed some light on the ghost star product and its different versions (reduced and
   twisted)
    which have been used in the algebraic and conformal methods.\\
Moreover,  if for the matter the correspondence
between algebraic and conformal approach is quite clear and actually
relies on the  isomorphism between CFT fields and their Fourier
transformed oscillators, the correspondence in the ghost sector is
more subtle since it compares, on the one hand, projectors squeezed states in
Siegel gauge build up with oscillators of the $bc$ system \cite{HKw} and, on the other hand, projectors
 surface states in a $bc$ CFT, twisted by one unit of
ghost current \cite{GRSZ1}; in this way the star product is ghost number preserving and we can properly
 define projectors.\\
 Since, when restricted to  Siegel gauge, we can define the reduced star
product which is also ghost number preserving, it was then natural to
identify the reduced star product on the algebraic side with the
twisted star product on the BCFT side. This is actually not correct\footnote{We thank L. Rastelli for
pointing out this.}, as we are going to show, since the
two product are different (they have different Neumann coefficients) but, curiously enough, they define
the same sliver-projectors.\\
The picture which arise at the
end of our analysis is that of three star products (matter, ghost and
twisted ghost); each of them can be defined completely from the
underlying CFT and all sliver--like projectors have the same (up to a minus
sign) Neumann
  coefficients. It is interesting  to
note that such equality of solutions is implied in a bijective way by the Gross-Jevicki relation
\cite{GJ2} which connects the ghost Neumann coefficients to the matter ones. Moreover the relevant
structures of the matter star product (at least at zero momentum) are  completely encoded in the
Neumann coefficients of the twisted $bc$ system. This fact
puts the $bc$ CFT (with its twisted variant) in an equivalent position with respect to the matter.\\
The paper is organized as follows. In section 2 we briefly recall definitions and properties of
the matter and ghost product from a  conformal point of view, as
done in \cite{tope}, then we perform the twist as in \cite{GRSZ1} and
define properly the twisted star product, which turns out to be very
closely related to the matter product.\\
In section 3 we determine the relations that connect the three vertices in the game.
In particular we show that the twisted CFT defines, up to a sign, the same coefficients as the matter.\\
In section 4, after a brief review of matter and ghost algebraic projectors, we derive the
algebraic expression of the twisted sliver and identify it with the
sliver--like state  in Siegel gauge, through the Gross--Jevicki relation
\cite{GJ2}.\\
 Section 5 is devoted to spectral analysis of the 2 star
products in the ghost sector: one  of them can be naturally casted in a matrix
structure which exhibits a discrete spectrum
consisting of one eigenvalue and one eigenvector (a $c$ midpoint
insertion) plus a continuous spectrum. Here the use of the twisted CFT
allows us to derive the continuous spectrum of the twisted product
following the same conformal methods of \cite{RSZ5} and equality of
twisted and non twisted slivers leads  to the spectrum of the
reduced product.
\footnote{On the same day the paper \cite{Kling} appeared,
with some overlapping with our results.}

%%%%%%%%%%%%%%%%%%%%%%%%%%%%%%%%%%%%%%%%%%%%%%%
\section{The three stars}
%%%%%%%%%%%%%%%%%%%%%%%%%%%%%%%%%%%%%%%%%%%%%%%
In this section we briefly review the construction of the interaction vertex
of matter and ghost sector. Details will be almost skipped and can be found in \cite{tope}
%%%%%%%%%%%%%%%%%%%%%%%%%%%%%
\subsection{Matter star}
%%%%%%%%%%%%%%%%%%%%%%%%%%%%%
The matter part of the three strings vertex \cite{W1,GJ1,GJ2} is given by
\beq
|V_3\rangle= \int d^{26}p_{(1)}d^{26}p_{(2)}d^{26}p_{(3)}
\delta^{26}(p_{(1)}+p_{(2)}+p_{(3)})\,{\rm exp}(-E)\,
|0,p\rangle_{123}\label{V3}
\eeq
where
\beq
E= \sum_{a,b=1}^3\left(\frac 12 \sum_{m,n\geq 1}\eta_{\mu\nu}
a_m^{(a)\mu\dagger}V_{mn}^{ab}
a_n^{(b)\nu\dagger} + \sum_{n\geq 1}\eta_{\mu\nu}p_{(a)}^{\mu}
V_{0n}^{ab}
a_n^{(b)\nu\dagger} +\frac 12 \eta_{\mu\nu}p_{(a)}^{\mu}V_{00}^{ab}
p_{(b)}^\nu\right) \label{E}
\eeq
Summation over the Lorentz indices $\mu,\nu=0,\ldots,25$
is understood and $\eta$ denotes the flat Lorentz metric.
The operators $ a_m^{(a)\mu},a_m^{(a)\mu\dagger}$ denote the non--zero
modes matter oscillators of the $a$--th string, which satisfy
\beq
[a_m^{(a)\mu},a_n^{(b)\nu\dagger}]=
\eta^{\mu\nu}\delta_{mn}\delta^{ab},
\quad\quad m,n\geq 1 \label{CCR}
\eeq
$p_{(r)}$ is the momentum of the $a$--th string and
$|0,p\rangle_{123}\equiv |p_{(1)}\rangle\otimes
|p_{(2)}\rangle\otimes |p_{(3)}\rangle$ is
the tensor product of the Fock vacuum
states relative to the three strings. $|p_{(a)}\rangle$ is
annihilated by
the annihilation
operators $a_m^{(a)\mu}$ and it is eigenstate of the momentum operator
$\hat p_{(a)}^\mu$
with eigenvalue $p_{(a)}^\mu$. The normalization is
\beq
\langle p_{(a)}|\, p'_{(b)}\rangle = \delta_{ab}\delta^{26}(p+p')\0
\eeq
The conformal definition of the vertex starts with  the gluing functions
\beq
f_a(z_a)=\alpha^{2-a} f(z_a) \, ,\, a=1,2,3
\eeq
where
\bea
f(z)&=&\Big{(} \frac{1+iz}{1-iz}\Big{)} ^{\frac{2}{3}}\\
\alpha&=&e^{\frac{2\pi i}{3}}\0
\eea
The interaction vertex is defined by a correlation function on the disk in
the following way
\beq\label{cover}
\int\psi*\phi*\chi=\langle f_1\circ\psi(0)\, f_2\circ\phi(0)\, f_3\circ\chi(0)\rangle=
\langle V_{3}|\psi\rangle_{1}|\phi\rangle_2|\chi\rangle_3
\eeq
Now we consider the string propagator
at two generic points of this disk. The Neumann coefficients $N^{ab}_{nm}$
are nothing but the Fourier modes of the propagator with respect to the
original coordinates $z_a$. We shall see that such Neumann coefficients
are related in a simple way to the standard three strings vertex coefficients.
Here we will deal only with the zero momentum vertex, which is the one which is strictly connected to the
(twisted) ghost vertex.

The Neumann coefficients $N_{mn}^{ab}$ are given by \cite{leclair1}
\beq
N_{mn}^{ab}=\langle V_3|\a_{-n}^{(a)}\a_{-m}^{(b)}|0\rangle_{123}=
-\frac{1}{nm}\oint\frac{dz}{2\pi i}\oint\frac{dw}{2\pi i}
\frac{1}{z^{n}}\frac{1}{w^{m}}f'_a(z)
\frac{1}{(f_a(z)-f_b(w))^2}f'_b(w)\label{neumann}
\eeq
where the contour integrals are understood around the origin.
It is easy to check that
\bea
N_{mn}^{ab}&=&N_{nm}^{ba}\nonumber\\
N_{mn}^{ab}&=&(-1)^{n+m}N_{mn}^{ba}\label{cycl}\\
N_{mn}^{ab}&=&N_{mn}^{a+1,b+1}\0
\eea
In the last equation the upper indices are defined mod 3.
\par
Let us consider the decomposition
\beq
N_{nm}^{ab}=\frac{1}{3\sqrt{nm}}{\Big{(}}C_{nm}+\bar{\a}^{a-b}U_{nm}+
\a^{a-b}\bar{U}_{nm}{\Big{)}}\label{decomp}
\eeq
After some algebra one gets
\bea
C_{nm}\!&=&\!\frac{-1}{\sqrt{nm}}\oint\frac{dz}{2\pi i}
\oint\frac{dw}{2\pi i}
\frac{1}{z^{n}}\frac{1}{w^{m}}
\Big{(}\frac{1}{(1+zw)^2}+\frac{1}{(z-w)^2}\Big{)}\\
\label{Anm}
U_{nm}\!&=&\!\frac{-1}{3\sqrt{nm}}\oint\frac{dz}{2\pi i}\oint\frac{dw}{2\pi i}
\frac{1}{z^{n}}\frac{1}{w^{m}}
\Big{(}\frac{f^2(w)}{f^2(z)}+2\frac{f(z)}{f(w)}\Big{)}
\Big{(}\frac{1}{(1+zw)^2}+\frac{1}{(z-w)^2}\Big{)} \0\\
\bar{U}_{nm}\!&=&\!\frac{-1}{3\sqrt{nm}}\oint\frac{dz}{2\pi i}
\oint\frac{dw}{2\pi i}
\frac{1}{z^{n}}\frac{1}{w^{m}} \Big{(}\frac{f^2(z)}{f^2(w)}+
2\frac{f(w)}{f(z)}\Big{)}
\Big{(}\frac{1}{(1+zw)^2}+\frac{1}{(z-w)^2}\Big{)} \0
\eea
The integrals can be directly computed in terms of the Taylor coefficients
of $f$. The result is
\bea
C_{nm}&=&(-1)^n\delta_{nm}\label{Anm1}\\
U_{nm}&=&\frac{1}{3\sqrt{nm}}\sum_{l=1}^{m}l\Big{[}(-1)^n
B_{n-l}B_{m-l}+2b_{n-l}b_{m-l}(-1)^m\0\\
       &&-(-1)^{n+l} B_{n+l}B_{m-l}-2b_{n+l}b_{m-l}(-1)^{m+l}\Big{]}\label{Unm1}\\
\bar{U}_{nm}&=&(-1)^{n+m}U_{nm}\label{Ubarnm1}
\eea
where we have set
\bea
f(z)&=&\sum_{k=0}^{\infty}b_kz^k\nonumber\\
f^2(z)&=&\sum_{k=0}^{\infty}B_kz^k, \quad \quad {\rm i.e.}\quad\quad
B_k = \sum_{p=0}^k b_p b_{k-p} \label{an}
\eea
Using the integral representation (\ref{Anm}) one can prove \cite{tope}
\beq\label{v1}
\sum_{k=1}^{\infty} U_{nk}U_{km}=\delta_{nm},\quad\quad
\sum_{k=1}^{\infty} \bar U_{nk} \bar U_{km}=\delta_{nm}\label{UU}
\eeq
In order to make contact with the standard notations (for example \cite{RSZ1}) we define
\footnote{The factor $(-1)^{n+m}$ is there because these coefficients refer to the Ket vertex
$|V_3\rangle$, so $bpz$ is needed.}
\beq
V_{nm}^{ab} = (-1)^{n+m}\sqrt{nm}\, N_{nm}^{ab}\label{identif1}
\eeq
and
\bea
M&=&CV^{11}\0\\
M_+&=&CV^{12}\\
M_-&=&CV^{21}\0
\eea
Using (\ref{UU}), together with the
decomposition (\ref{decomp}), it is easy to establish the following linear and non linear
relations (written in matrix notation).
\bea
&M+M_++M_-=1&\0\\
&M^2+M_+^2+M_-^2=1&\0\\
&M_+^3+M_-^3=2M^3-3M^2+1&\0\\
&M_+M_-=M^2-M&\label{mprop}\\
&[M,M_{\pm}]=0&\0\\
&[M_+,M_-]=0&\0
\eea

%%%%%%%%%%%%%%%%%%%%%%%%%%%%%%%%%%
\subsection{Ghost star}
%%%%%%%%%%%%%%%%%%%%%%%%%%%%%%%%%%%

To start with we define, in the ghost sector, the vacuum states $|\hat
0\rangle$ and $|\dot 0\rangle$ as follows
\beq |\hat 0\rangle =
c_0c_1|0\rangle, \quad\quad |\dot 0\rangle = c_1|0\rangle
\label{vacua}
\eeq
where $|0\rangle$ is the usual $SL(2,\RR)$ invariant vacuum.
Using $bpz$ conjugation
\bea c_n\rightarrow
(-1)^{n+1}c_{-n},\quad\quad b_n\rightarrow (-1)^{n-2}b_{-n},
\quad\quad |0\rangle \rightarrow \langle 0|
\eea
one can define
conjugate states.

The three strings interaction vertex is defined, as usual, as a
squeezed operator acting on three copies of the $bc$ Hilbert space
\beq
\langle \tilde V_{3}|=\, _1\!\langle\hat{0}|\,
_2\!\langle\hat{0}|\, _3\!\langle\hat{0}|e^\tE, \quad\quad
\tE=\sum_{a,b=1}^3\sum_{n,m}^{\infty}c_n^{(a)}\tN_{nm}^{ab}b_m^{(b)}
\label{V3gh}
\eeq
Under $bpz$ conjugation
\beq
|\tilde
V_{3}\rangle=e^{\tE'}|\hat{0}\rangle_1|\hat{0}\rangle_2|\hat{0}
\rangle_3,\quad\quad
{\tE'}=-\sum_{a,b=1}^3\sum^\infty_{n,m}(-1)^{n+m}c_n^{(a)\,
\dagger}\tN_{nm}^{ab}b_m^{(b)\, \dagger}\label{V3gh'}
\eeq
To make the propagator $SL(2,\RR)$ we have to insert three $c$ zero modes at points $\xi_i$  \cite{leclair1}
\beq
\langle b(z)
c(w)\rangle=\frac{1}{z-w}\prod_{i=1}^3 \frac{w-\zeta_i}{z-\zeta_i}
\eeq
So we get
\bea
\tN_{nm}^{ab}&=&\langle
\tilde{V}_3|b_{-n}^{(a)}c_{-m}^{(b)}|\dot{0}\rangle_{123\label{Nnmtil}}\\
&=&\oint\frac{dz}{2\pi i}\oint\frac{dw}{2\pi i}\frac{1}{z^{n-1}}
\frac{1}{w^{m+2}}(f'_a(z))^2
\frac{-1}{f_a(z)-f_b(w)}\prod_{i=1}^3\frac{f_b(w)-f_i(0)}{f_a(z)-f_i(0)}
(f'_b(w))^{-1}\0
\eea
It is straightforward to check that
\beq
\tN_{nm}^{ab}=\tN_{nm}^{a+1,b+1}\label{cyclgh}
\eeq and (by letting
$z\rightarrow -z,\, w\rightarrow -w$)
\beq
\tN_{nm}^{ab}=(-1)^{n+m}\tN_{nm}^{ba}\label{twistgh}
\eeq
As in the matter case, we consider the decomposition
\beq
\tN_{nm}^{ab}=\frac{1}{3}(\tC_{nm}+\bar{\alpha}^{a-b}\tU_{nm}+
\alpha^{a-b}\bar{\tU}_{nm})\label{decompgh}
\eeq
After some algebra one finds
\bea
\tC_{nm}&=&\oint\frac{dz}{2\pi i}\oint\frac{dw}{2\pi
i}\frac{1}{z^{n+1}} \frac{1}{w^{m+1}}
\Big{(}\frac{1}{1+zw}-\frac{w}{w-z}\Big{)}\nonumber\\
\label{U}
\tU_{nm}&=&\oint\frac{dz}{2\pi i}\oint\frac{dw}{2\pi
i}\frac{1}{z^{n+1}}\frac{1}{w^{m+1}}\frac{f(z)}{f(w)}
\Big{(}\frac{1}{1+zw} -\frac{w}{w-z}\Big{)}\\
\bar{\tU}_{nm}&=&\oint\frac{dz}{2\pi i}\oint\frac{dw}{2\pi
i}\frac{1}{z^{n+1}}\frac{1}{w^{m+1}}\frac{f(w)}{f(z)}
\Big{(}\frac{1}{1+zw} -\frac{w}{w-z}\Big{)}\nonumber
\eea
It is easy to show that
\beq
\bar{\tU}_{nm}  = (-1)^{n+m}{\tU}_{nm} \label{UUbar}
\eeq
As discussed in detail in \cite{tope} the evaluation of these integrals is sensible to
 radial ordering in the $(n,-n)$, components. We fix the ambiguity by setting
\beq
\tN^{aa}_{-1,1} = \tN^{aa}_{1,-1}
=0, \quad \quad \tN^{aa}_{0,0} =1.
\label{ambfix}
\eeq
Which corresponds to
\bea
\tC_{NM}&=&(-1)^N\delta_{NM}\quad N,M\geq 0\\
\tU_{NM}&=&(-1)^Mb_Nb_M+(-1)^M\sum_{l=1}^M(b_{N-l}b_{M-l}+ (-1)^l
b_{N+l}b_{M-l})
\eea
where the $b_n$'s have been defined in (\ref{an}).
The reason for this is that we get the fundamental identity
\beq
\sum_{K=0}
\tU_{NK}\tU_{KM} = \delta_{NM}\label{UU'}
\eeq
As for the matter case we will consider from now on the coefficients of  the Ket vertex
\beq
\V_{NM}^{ab}=-(-1)^{n+m}\tN_{NM}^{ab}
\eeq
Then we define
\bea
\Y&=&C\V^{11}\0\\
\Y_+&=&C\V^{12}\0\\
\Y_-&=&C\V^{21}\0
\eea
From (\ref{UU'}) we obtain the linear and non linear relations
\bea
&\Y+\Y_++\Y_-=1&\0 \\
&\Y^2+\Y_+^2+\Y_-^2=1&\0\\
&\Y_+^3+\Y_-^3=2\Y^3-3\Y^2+1&\0\\
&\Y_+\Y_-=\Y^2-\Y&\label{bprop}\\
&[\Y,\Y_{\pm}]=0&\0\\
&[\Y_+,\Y_-]=0&\0
\eea
Note that the matrices $\Y$ have the structure
\bea\label{struct}
\Y&=&\left(\matrix{1& 0\cr \vec{y}& Y}\right)\\
\Y_\pm&=&\left(\matrix{0& 0\cr \vec{y}_\pm & Y_\pm}\right)
\eea
The ``small'' matrices $Y,Y_\pm$ are the coefficients of the reduced star product $*_{b_0}$, defined as
\beq
A*_{b_0}B=b_0(A*B)
\eeq
From (\ref{struct}) the  following properties of ``small'' matrices and  column vectors relative to the $b_0$ modes (hence excluded from the reduced product) are inherited
\bea
&Y+Y_++Y_-=1\nonumber\\
 &\vec{y}+\vec{y}_++\vec{y}_-=0&\nonumber\\
&Y^2+Y_+^2+Y_-^2=1&\nonumber\\
&(1+Y)\vec{y}+Y_+\vec{y}_++Y_-\vec{y}_-=0&\0\\
&Y_+^3+Y_-^3=2Y^3-3Y^2+1&\0\\
&Y_+^2\vec{y}_+ +Y_-^2\vec{y}_-
=(2Y^2-Y-1)\vec{y}&\label{lprop}\\
&Y_+Y_-=Y^2-Y&\0\\
&[Y,Y_{\pm}]=0&\nonumber\\
&[Y_+,Y_-]=0&\nonumber\\
&Y_+\vec{y}_-=Y\vec{y}= Y_-\vec{y}_+&\nonumber\\
&-Y_{\pm}\vec{y}=(1-Y)\vec{y}_{\pm}&\nonumber
\eea
To end this paragraph we mention the important fact that in order to use big matrices
in practical computation, one has to enlarge the zero mode sector, in order to disentangle
 the non-anticommuting oscillators $(b_0,c_0)$. This technique is explained in detail in \cite{tope}.

%%%%%%%%%%%%%%%%%%%%%%%%%%%%%%%%%%%%
\subsection{The twisted star}
%%%%%%%%%%%%%%%%%%%%%%%%%%%%%%%%%%%%%%

In \cite{GRSZ1} another type of star-product is considered. It represents the gluing condition
in a  twisted conformal field theory of the ghost system. The twist is done by subtracting to the
stress tensor one unit of derivative of the ghost current
\beq
\label{twist} T'(z)=T(z)-\partial{j_{gh}(z)}
\eeq
This redefinition changes the conformal weight of the $bc$ fields from (2,-1) to (1,0). It follows
 that the background charge is shifted from -3 to -1. As a consequence, in order not to have vanishing
 correlation functions, we have to fix only one $c$ zero-mode. In particular,
the $SL(2,\RR)$--invariant
   propagator of the $bc$ system is
\beq
\langle b(z)c(w)\rangle '=\frac{1}{z-w}\frac{w-\xi}{z-\xi}
\eeq
where $\xi$ is one fixed point.\\
In \cite{GRSZ1} it was shown that the usual product can be obtained from the twisted one by
inserting a $n_{gh}=1$--operator at the midpoint which, on singular states like the sliver,
can be identified with a $c$--midpoint insertion. This implies that, on such singular projectors,
the twisted product can be identified with the reduced one.\\
The twisted ghost Neumann coefficients are then defined to be\footnote{We put, for simplicity, $\xi=f_a(i)=0$}
\bea
\tN_{nm}'^{ab}=\oint\frac{dz}{2\pi i}\oint\frac{dw}{2\pi i}
\frac{1}{z^{n}}\frac{1}{w^{m+1}}f'_a(z)
\frac{-1}{f_a(z)-f_b(w)}
\frac{f_b(w)}{f_a(z)}\0\\
=\oint\frac{dz}{2\pi i}\oint\frac{dw}{2\pi i}
\frac{1}{z^{n}}\frac{1}{w^{m+1}}
\frac{4i}{3}\frac{1}{1+z^2}
\frac{\bar{\alpha}^{b}f(w)}{\bar{\alpha}^{a}f(z)-\bar{\alpha}^{b}f(w)}
\eea
As in (\ref{Nnmtil}) these coefficients refer to the Bra vertex, the corresponding coefficients for the Ket vertex are
\beq
\tV_{nm}'^{ab}=-(-1)^{n+m}\tN_{nm}'^{ab}
\eeq
We will see in the next section how to compute such coefficients using previous results. This will
lead to interesting connections with the other star-products.
%%%%%%%%%%%%%%%%%%%%%%%%%%%%%%%%%%%%%%%%%%%%%%%%
\section{Relations among the stars}
%%%%%%%%%%%%%%%%%%%%%%%%%%%%%%%%%%%%%%%%%%%%%%%
In this section we will show how the stars products defined above are related  to each other. In
 particular we will show the explicit relations which connect all the Neumann coefficients in
 the game, so at the end the three star products are homeomorphic and in this sense can be considered equivalent.

%%%%%%%%%%%%%%%%%%%%%%%%%%%%%%%%%%%%%%%
\subsection{Twisted ghosts vs Matter}
%%%%%%%%%%%%%%%%%%%%%%%%%%%%%%%%%%%%%%%

The commuting matter Neumann coefficients which appear in (\ref{mprop})  are given by
\beq
M_{nm}^{ab}=
-\frac{(-1)^m}{\sqrt{nm}}\oint\frac{dz}{2\pi i}\oint\frac{dw}{2\pi i}
\frac{1}{z^{n}}\frac{1}{w^{m}}f'_a(z)
\frac{1}{(f_a(z)-f_b(w))^2}f'_b(w)
\eeq
We can  rewrite them  as
\bea
M_{nm}^{ab}= -\frac{(-1)^m}{\sqrt{nm}}\oint\frac{dz}{2\pi i}\oint\frac{dw}{2\pi i}
\frac{1}{z^{n}}\frac{1}{w^{m}}f'_a(z) \d_w
\frac{1}{f_a(z)-f_b(w)}\nonumber\\
=-(-1)^{m}{\sqrt{\frac{m}{n}}}\oint\frac{dz}{2\pi i}\oint\frac{dw}{2\pi i}
\frac{1}{z^{n}}\frac{1}{w^{m+1}}
\frac{f'_a(z)}{f_a(z)-f_b(w)}
\eea
where we have integrated by part to respect the variable $w$. Now, recalling
\beq
f'_a(z)=\frac{4i}{3}\frac{1}{1+z^2}\alpha^{2-a}f(z)\ ,
\eeq
we obtain
\beq
M_{nm}^{ab}=-(-1)^m{\sqrt{\frac{m}{n}}}\oint\frac{dz}{2\pi i}\oint\frac{dw}{2\pi i}
\frac{1}{z^{n}}\frac{1}{w^{m+1}}
\frac{4i}{3}\frac{1}{1+z^2}
\frac{\bar{\alpha}^{a}f(z)}{\bar{\alpha}^{a}f(z)-\bar{\alpha}^{b}f(w)}
\eeq
Let us now consider the corresponding twisted ghost Neumann coefficients
\bea
Y_{nm}'^{ab}&=&(C\tV'^{ab})_{nm}\0\\
&=&(-1)^{m}\oint\frac{dz}{2\pi i}\oint\frac{dw}{2\pi i}
\frac{1}{z^{n}}\frac{1}{w^{m+1}}
\frac{f'_a(z)}{(f_a(z)-f_b(w))}
\frac{f_b(w)}{f_a(z)}\0\\
&=&(-1)^{m}\oint\frac{dz}{2\pi i}\oint\frac{dw}{2\pi i}
\frac{1}{z^{n}}\frac{1}{w^{m+1}}
\frac{4i}{3}\frac{1}{1+z^2}
\frac{\bar{\alpha}^{b}f(w)}{(\bar{\alpha}^{a}f(z)-\bar{\alpha}^{b}f(w))}
\eea
This coefficients are not symmetric if we exchange $n$ with $m$, however we can easily symmetrize
them by the use of the matrix $E_{nm}=\sqrt{n}\delta_{nm}$
\beq
Y'^{ab}\rightarrow E^{-1}Y'^{ab}E
\eeq
It is now easy to show the following
\bea
(E^{-1}Y'^{ab}E)_{nm}+M_{nm}^{ab}
&=&(-1)^{m}{\sqrt{\frac{m}{n}}}\oint\frac{dz}{2\pi i}\oint\frac{dw}{2\pi i}
\frac{1}{z^{n}}\frac{1}{w^{m+1}}
\frac{4i}{3}\frac{1}{1+z^2}
\frac{(\bar{\alpha}^{b}f(w)-\bar{\alpha}^{a}f(z))}{(\bar{\alpha}^{a}f(z)-\bar{\alpha}^{b}f(w))}\0\\
&=&-(-1)^{m}{\sqrt{\frac{m}{n}}}\oint\frac{dz}{2\pi i}\oint\frac{dw}{2\pi i}
\frac{1}{z^{n}}\frac{1}{w^{m+1}}
\frac{4i}{3}\frac{1}{1+z^2}=0
\eea
 the last equality holding since there are no poles for $n,m\geq 1$.\\
So we obtain
\beq
E^{-1}Y'^{ab}E=-M^{ab}
\eeq
a remarkable relation between twisted ghost and matter vertices, which is the same  relation
 that holds in the four-string vertex between the non-twisted ghost and the matter Neumann coefficients \cite{GJ2}.
This relation proves also  that the ghost integral is independent of the background charge,
 for $n,m\geq 1$: the matter integral, indeed, can be seen as the ghost integral without the
  background charge\footnote{The independence of the background charge is also crucial to prove $\tN'^{ab}=C\tN'^{ba}C$}.
As a consequence of the relation with the matter coefficients we can derive all the relevant
 properties of the twisted ghost Neumann coefficients, by simply taking the
matter results (\ref{mprop}) and
  changing the sign in odd powers.
\bea
&Y'+Y'_++Y'_-=-1&\nonumber\\
&Y'^2+Y_+'^2+Y_-'^2=1&\nonumber\\
&Y_+'^3+Y_-'^3=2Y'^3+3Y'^2-1&\0\\
&Y'_+Y'_-=Y'^2+Y'&\label{tprop}\\
&[Y',Y'_{\pm}]=0&\nonumber\\
&[Y'_+,Y'_-]=0&\nonumber
\eea

%%%%%%%%%%%%%%%%%%%%%%%%%%%%%%%%%%%%%%%%
\subsection{Twisted vs Reduced }
%%%%%%%%%%%%%%%%%%%%%%%%%%%%%%%%%%%%%%%%
The relation between the twisted and non-twisted ghost Neumann coefficients
can now be
 obtained using the previous relation
\beq
Y'=-EME^{-1}\label{tm}
\eeq
and the Gross-Jevicki relation \cite{GJ2}\footnote{This relation, as noted in \cite{belov} contains the map
\beq
P(z)= \frac{-z}{1+2z}
\eeq
which is a PSL(2,R) transformation that squares to itself
\beq
P\circ P(z)=z
\eeq
}
\beq\label{gjr}
Y=E\frac{-M}{1+2M}E^{-1}
\eeq
 between matter and non-twisted ghosts.
So, finally, we have
\beq
Y=\frac{Y'}{1-2Y'}\label{tjr}
\eeq
or
\beq
Y'=\frac{Y}{1+2Y}
\eeq

This relation is also strictly related to the equality of solutions between the ghost
sliver constructed from the twisted CFT and the non-twisted one \cite{okuda}. Indeed,
it is possible to derive such relation from the equality of ghost algebraic  slivers,
 as we will see in the next section.

%%%%%%%%%%%%%%%%%%%%%%%%%%%%%%%%%%%%%%%%%%%%%%%
\section{Slivers}
%%%%%%%%%%%%%%%%%%%%%%%%%%%%%%%%%%%%%%%%%%%%%%%
In this section we review the algebraic derivation of the sliver state in matter and
ghost sector. Then we compute algebraically the slivers in the twisted ghost sector
and show how identity of such states is implied by the relations (\ref{tjr}) between
the Neumann coefficients in the game.
%%%%%%%%%%%%%%%%%%%%%%%%%%%%%%%%%%%%%%%%%%%%%%%
\subsection{Matter sliver and Ghost solution}
%%%%%%%%%%%%%%%%%%%%%%%%%%%%%%%%%%%%%%%%%%%%%%%
The projection equation  in the matter sector
\beq
\label{me}
|\psi\rangle_m=|\psi\rangle_m\ast_m|\psi\rangle_m
\eeq
can be solved as in \cite{KP,RSZ2}, by the ansatz
\bea
|\psi\rangle_m &=& \N_m\exp\left(\sum_{n,m\geq 1}
a_n^{\dagger}S_{nm}a_{m}^{\dagger}\right)|0\rangle\\
S&=&CSC
\eea
where
\beq
T=CS=\frac{1}{2M}\left(1+M-\sqrt{(1-M)(1+3M)}\right)
\eeq
The ghost equation of motion is
\beq
\label{ge}
\Q|\psi\rangle_g+|\psi\rangle_g\ast_g|\psi\rangle_g=0
\eeq
This equation is easy to solve if we  use big matrices in order to handle at the
same time both zero and non zero modes (see \cite{tope} ). The relevant results are
\bea
|\psi\rangle_g &=&\tilde\N_g
\exp\left(\sum_{n,m\geq 1}c_n^{\dagger}\tS_{nm}b_{m}^{\dagger}\right)|\dot{0}\rangle\label{Sdot}\\
\tT&=&C\tS=\frac{1}{2Y}{\Big{(}}1+Y-\sqrt{(1-Y)(1+3Y)}{\Big{)}}\\
\Q&=& c_0 + \vec f\cdot(\vec c +C\vec c^{\dagger})\label{trueQ}\\
\vec f&=&\frac{\vec y}{1-Y}
\eea
Using the integral representations (\ref{U}) one can actually prove that $\Q$ is a midpoint insertion \cite{tope, Oku2}
\beq\label{sol}
\Q = c_0 + \sum_{n=1}^\infty (-1)^n(c_{2n} +c_{-2n})=\frac{1}{2i}\left(c(i)-c(-i)\right)
\eeq

%%%%%%%%%%%%%%%%%%%%%%%%%%%%%%%%%%%%%%%%%%%%%%%%%%%%%%%%%%
\subsection{The twisted sliver in the algebraic approach}
%%%%%%%%%%%%%%%%%%%%%%%%%%%%%%%%%%%%%%%%%%%%%%%%%%%%%%%%%%

We have seen that the Neumann coefficients of the star product in the
twisted CFT coincides to (minus) the matter ones at zero
momentum. This implies that we can solve the algebraic equation for
projectors, as for the usual ghost star product, but now using the
linear and non linear relations (\ref{tprop}).\\
So we impose the projector equation
\beq
|S\rangle'=|S\rangle'*'|S\rangle'
\eeq
with the ansatz
\beq |S\rangle' =\N \exp\left(\sum_{n,m\geq
1}c_n^\dagger S'_{nm}b_{m}^\dagger \right)|0'\rangle\label{S'}
\eeq
we can safely follow  the non twisted case \cite{HKw, tope} and  arrive at the equation

\bea
T'=CS'= Y' +(Y'_+,Y'_-)\frac 1{ 1-\Sigma'{{\cal V'}}}\Sigma'
\left(\matrix{Y'_-\cr Y'_+}\right)\label{S2'} \eea where \bea \Sigma'=
\left(\matrix{T'&0\cr 0& T'}\right), \quad\quad\quad {\cal V'} =
\left(\matrix{Y'&Y'_+\cr Y'_-&Y'}\right).\0
\eea
using the properties
(\ref{prop'}) we obtain the algebraic equation
\beq
(T'+1)(-Y' T'^2 + (1-Y')T' -Y' )=0
\eeq
which, apart from the trivial solution $T'=-1$, gives\footnote{As usual we
choose the square root branch cut which doesn't have divergence as $Y'\rightarrow 0$.}

\beq\label{sol'}
T'=CS'=\frac{1}{2Y'}{\Big{(}}1-Y'-\sqrt{(1+Y')(1-3Y')}{\Big{)}}
\eeq
It is interesting to compare it with the algebraic projector
w.r.t. the reduced product
\beq
\tT=\frac{1}{2Y}{\Big{(}}1+Y-\sqrt{(1-Y)(1+3Y)}{\Big{)}}
\eeq
The equality of this two solutions holds if and only if the following
relation between twisted and non twisted Neumann coefficients is
obeyed
\beq
Y'=\frac{Y}{1+2Y}
\eeq
which is exactly (\ref{tjr}).
This shows that equality of solution in VSFT is equivalent to the statement
(\ref{gjr}) which, on the other hand, have its explanation via the
4-string vertex \cite{GJ2}. It would  be interesting then to explore
further the relations (if any) between matter and ghost in the
4--string vertex and matter and twisted ghost in the
conformal--three--string vertex.
%%%%%%%%%%%%%%%%%%%%%%%%%%%%%%%%%%%%%%%%%
\section{Diagonalization}
%%%%%%%%%%%%%%%%%%%%%%%%%%%%%%%%%%%%%%%%%
In this section we compute the spectrum of the twisted product and verify
 that is related to the matter one by (\ref{tm}). Then we do the same thing
  on the usual ghost product. In this case the diagonalization of ``big'' matrices can
be carried over in two step. First we block diagonalize them isolating
completely the (0,0) component: this will determine a sort of discrete
spectrum. Second we diagonalize the internal (small) matrices, using the results from the twisted CFT,
in order to find the usual continuous spectrum.\footnote{The topic of ghost spectroscopy in Siegel gauge
 was also treated in \cite{Erler, Barso, Belov8}}

%%%%%%%%%%%%%%%%%%%%%%%%%%%%%%%%%%%%%%%%%
\subsection{Diagonalization of  the twisted product}
%%%%%%%%%%%%%%%%%%%%%%%%%%%%%%%%%%%%%%%%%

Knowing the fact that twisted Neumann coefficients can be easily
symmetrized to take the form of (minus) the matter Neumann
coefficients, we have for free the eigenvalues. However we would like
to show, as a consistency check, that we can derive the twisted
spectrum by purely conformal considerations, following the lines of
\cite{RSZ5} but now using  the twisted conformal field theory of the
ghost system. As we have seen before, the twist is done as
\beq
 T'(z)=T(z)-\partial{j_{gh}(z)}
\eeq
leading to
\beq\label{viras'}
L_n'=L_n+nj_n+\delta_{n0}
\eeq
where
\bea L_n&=&-\sum_{k=-\infty}^{\infty}(2n-k):c_{n-k}b_k:\\
j_n&=&\sum_{k=-\infty}^{\infty}:c_{n-k}b_k:\0
\eea
To find the eigenvectors of $Y$ we  consider the $*'$ algebra
derivation
\beq\label{K1}
K_1'=L_{1}'+L_{-1}'
\eeq
and then we use the
same formal arguments of \cite{RSZ5}.  The main difference
here is that $K_1'$ acts on $b$ and $c$ oscillator in a
different but complementary way, due to their (twisted) conformal
properties
\beq
[K_1',c_n]=-(n+1)c_{n+1}-(n-1)c_{n-1}\0
\eeq
\beq
[K_1',b_n]=-n\, b_{n+1}-n\, b_{n-1}\0
\eeq
We can have $c$-type vectors $v_n$, as well as $b$-type vectors $w_n$, so $K_1$ has two
different matrix representations.  If we act on $c$ oscillators we get
\bea
[K_1',v_nc_n]&=&(K^{(c)}v)\cdot c\0\\
K^{(c)}_{nm}&=&-(m+1)\delta_{n,m+1}-(m-1)\delta_{n,m-1}
\eea
If we act on $b$ oscillators we get
\bea
[K_1',w_nb_n]&=&(K^{(b)}v)\cdot
b\0-w_1b_0\\ K^{(b)}_{nm}&=&-m\delta_{n,m+1}-m\delta_{n,m-1}
\eea
These two  matrices transpose to each other and obey
\beq\label{trans}
K^{(c)}=K^{(b)\, T}=A^{-1}K^{(b)}A
\eeq
in particular they share
eigenvalues. The matrix $A$ is defined to be
\beq
A_{nm}=n\delta_{nm}
\eeq
We shall begin by diagonalizing $K^{(c)}$ and determine its
eigenvectors.
\beq\label{eigc}
K^{(c)}\, v^{\k}=\k\, v^{\k}
\eeq
In order to do so we map this algebraic problem in a differential one, by
defining the generating function
\beq f_{v^{\k}}(z)=\sum_{n=1}^{\infty}v_n^{\k}z^n
\eeq
so that
\beq
v_n^{\k}=\oint_0 \frac{dz}{2\pi i}\frac{1}{z^{n+1}}f_{v^{\k}}(z)
\eeq
With trivial manipulations we find that (\ref{eigc}) is equivalent to
\beq
\left(-(1+z^2)\frac{d}{dz}-(z-\frac 1z)\right)f_{v^{\k}}(z)=\k\,
f_{v^{\k}}(z)
\eeq
which is easily integrated to give
\beq
f_{v^{\k}}(z)=\frac{z}{z^2+1}\e^{-\k \tan^{-1}z}
\eeq
where we have
chosen the overall normalization in order to $v_1^{\k}=1$.  As usual
$\k$ is a continuous parameter spanning all the real axis.\\ To find
the $b$-eigenvectors it is worth noting that $K^{(b)}$ is the same as
in the matter case \cite{RSZ5}, so we simply get the result
\beq
f_{w^{\k}}(z)=\frac1k(1-\e^{-\k \tan^{-1}z})
\eeq
As a consistency
check note that due to (\ref{trans}) $c$-eigenvectors are related to
$b$-eigenvectors by

\beq
v_n^{\k}=n\, w_n^{\k}
\eeq
which in functional language reads
\beq
f_{v^{\k}}(z)=z\frac{d}{dz}f_{w^{\k}}(z)
\eeq
It is trivial to
check that this relation is identically satisfied.\\ Once the spectrum
of $K_1'$ is found, in order to find the spectrum of $Y$, we begin
by considering the algebra of wedge states in the twisted CFT. A wedge
state can be defined as

\beq\label{wedge'}
|N\rangle'=\left(|0\rangle'\right)_{*'}^{N-1}=
\N'_N \exp\left(\sum_{n,m=1}^\infty c_n^\dagger\, (CT'_N)_{nm}\,
b_m^\dagger\right)|0\rangle'
\eeq
These states satisfy the relation

\beq
|N+1\rangle' = |N\rangle' *' | 0\rangle'
\eeq
Following the same formal arguments of \cite{Furu}, we can write all $T_N$ in function of
the sliver matrix $T$\footnote{ Note the change of signs with respect
to \cite{Furu}, they come out from the differences in the algebraic
linear and non linear properties of the Neumann coefficients of the
twisted CFT}

\beq\label{wedg'}
T'_N=\frac{T'-T'^{N-1}}{1-T'^N}
\eeq
In particular we have
\bea
T'_2&=&0\\ T'_3&=&Y'\\ T'_\infty&=&T'
\eea
actually the last equation
is well defined for $|T|\leq1$, we will see a posteriori that the
eigenvalues of $T$ lie on the interval $(0,1]$.

Such wedge states can be defined as surface states in the twisted CFT
\cite{GRSZ1}. Given a string field $|\phi\rangle=\phi'(0)|
0\rangle'$ the wedge state $|N\rangle$ can be defined as\footnote{In
the brackets insertion of the $c_0$ 0 mode is intended, since all
oscillators in the game start from the 1 component, we don't have any
ambiguity}
\beq
'\langle N|\phi\rangle=\langle f_N\circ \phi(0)\rangle'
\eeq
where the generating function of the surface state is given by

\beq
f_N(z)=\frac N2 \tan\left( \frac 2N \tan^{-1} z\right)
\eeq
Now we consider the state $|2+\epsilon\rangle'$.  This state can be given
a representation in terms of the twisted Virasoro generators as
\cite{leclair2}

\bea
|B\rangle'&=&\exp\left(\epsilon V_-\right)|0\rangle'=|0\rangle'+\epsilon V'_-|0\rangle'+O(\epsilon^2)\\
V_-&=&\sum_{n=1}^{\infty}(-1)^n \frac{1}{(2n-1)(2n+1)} L'_{-2n} \label{V-}
\eea
Using the explicit form of the
twisted Virasoro generators
\beq
L'_n=-\sum_{k=-\infty}^{\infty}(n-k):c_{n-k}b_k:
\eeq
we can find the
relevant Neumann coefficients of the  state $|2+\epsilon\rangle'$
\beq\label{2+e}
|2+\epsilon\rangle'=\exp\left(\epsilon\sum_{n,m=1}^\infty
c_n^\dagger\, (CB')_{nm}\, b_m^\dagger\right)|0\rangle'
=|0\rangle'+\epsilon\sum_{n,m=1}^\infty c_n^\dagger\, (CB')_{nm}\,
b_m^\dagger|0\rangle'+O(\epsilon^2)
\eeq
the $B'_{nm}$ coefficients can be computed by comparing (\ref{2+e}) with (\ref{V-}), we get
\beq\label{B}
B'_{nm}=\frac 12 \left( 1+(-1)^{n+m}\right)
\frac{(-1)^{\frac{n-m}{2}} n}{(n+m)^2-1}
\eeq
This coefficient is made diagonal with $c$-type eigenvectors
\beq
\sum_{m=1}^{\infty}B'_{nm}v^{\k}_m =\sum_{m=1}^{\infty} B'_{nm} m\
w^{\k}_m=\beta'(\kappa) v^{\k}_n=\beta'(\kappa)n\ w^{\k}_n
\eeq
Now take
$n=1$, all goes the same way as \cite{RSZ5}, except for a minus sign
in the definition (\ref{B})
\beq
\beta'(\k)=\frac 12\
\frac{\frac{\pi\kappa}{2}}{\sinh\frac{\pi\kappa}{2}}
\eeq
From $B'$-eigenvalues we can find out the eigenvalues of the twisted
sliver $\tau'(k)$, by inverting the relation (\ref{wedg'}) at $N=2+\epsilon$
\beq
B'=\frac{T'\log(T')}{1-T'}
\eeq
which is bijective in the range
$T\in(0,1]$, in so doing we get
\beq\label{tau'}
\tau'(k)=\e^{-\frac{\pi|\kappa|}{2}}
\eeq
Then we can use the
twisted wedge states formula at $N=3$ to get the eigenvalues of $Y'$,
which we call $y'(\k)$
\beq
y'(\k)=\frac{\tau'(k)-\tau'(k)^2}{1-\tau'(k)^3}=\frac{1}{2\cosh
\frac{\pi\kappa}{2}+1}
\eeq

To find the spectrum of the other two coefficients $Y'_\pm$ we use the
relations

\bea\label{prop'}
Y' + Y'_+ + Y'_-&=&-1\0\\ Y'_+Y'_-&=&Y'^2+Y'\0
\eea
solving them for $Y'_\pm$ we get
\beq
Y'_\pm=-\frac 12 \left( 1+Y'\mp
\sqrt{(1-3Y')(1+Y')}\right)=-\frac{1+\cosh \frac{\pi\kappa}{2}\pm\sinh
\frac{\pi\kappa}{2}}{2\cosh \frac{\pi\kappa}{2}+1}
\eeq
As expected they are exactly the opposite of the matter ones

%%%%%%%%%%%%%%%%%%%%%%%%%%%%%%%%%%%%%
\subsection{Block diagonalization of non twisted star}
%%%%%%%%%%%%%%%%%%%%%%%%%%%%%%%%%%%%%

Let's rewrite for the sake of clarity the general form of the matrices defining the usual ghost product
\bea
\Y&=&\left(\matrix{1& 0\cr \vec{y}& Y}\right)\\
\Y_\pm&=&\left(\matrix{0& 0\cr \vec{y}_\pm & Y_\pm}\right)
\eea
The $(0,0)$ component isolates one eigenvalue for each matrix
\bea
eig[\Y]&=& 1 \oplus eig[Y]\\
eig[\Y_\pm]&=& 0 \oplus eig[Y_\pm]
\eea
It is then straightforward to find the eigenvector relative to these eigenvalues, this is achieved by block
 diagonalizing such matrices
\bea\label{blockdiag}
\hat\Y&=&\left(\matrix{1&
0\cr 0& Y}\right)\\
\hat\Y_\pm &=&\left(\matrix{0& 0\cr 0 &
Y_\pm}\right)
\eea
with the change of basis
\bea
\hat\Y_{(\pm)}&=&{\cal Z}^{-1}{\Y}_{(\pm)}{\cal Z}\\
{\cal Z}&=&\left(\matrix{1& 0\cr \vec{f}& 1}\right)\label{Z}\\
{\cal Z}^{-1}&=&\left(\matrix{1& 0\cr -\vec{f}& 1}\right)
\eea
where
\beq\label{f'}
\vec{f}=\frac{1}{1-Y}\vec{y}=-\frac{1}{Y_\pm}\vec{y}_\pm
\eeq
The equality of the last expressions is a simple consequence of (\ref{lprop}). Note that the
eigenvector we find is the same which defines the kinetic operator $Q$ as a $c$ midpoint insertion
\footnote{This, from a different point of view, was also note in \cite{Oku1}} .
Since the equation (\ref{f'}) has the solution (\ref{sol}), it might seem that $(Y,Y_\pm)$
(small matrices) cannot have the eigenvalues $(1,0)$, this is actually not true because
(\ref{f'}) is not a relation  in the full Hilbert space, but only in its twist-even subspace.
As we will see the $(1,0)$ eigenvalues will have a corresponding  one twist odd eigenvector,
contrary with all the other eigenvalues which will have eigenvectors of both twist parity.
The linear transformation (\ref{Z}) induces the following redefinition of the  $bc$ oscillators.
\bea\label{midp}
\tilde c_0&=&c_0 +\sum_{n\geq 1}f_n(c_n+(-1)^nc_n^\dagger)=\Q\\
\tilde c_n&=&c_n\quad n\neq 0\\
\tilde b_0&=&b_0\\
\tilde b_n&=&-f_nb_0+b_n\quad n\neq 0
\eea
where we have defined
$(f_{-n}\equiv f_n)$.  This is an
equivalent representation of the $bc$ system\footnote{In order to
prove this, twist invariance of $\vec{f}$ is crucial
($C\vec{f}=\vec{f}$)}
\beq
\{\tilde b_N,\tilde c_M\}=\delta_{N+M}\quad
N,M=-\infty,...,0,...,\infty
\eeq
Block diagonalization of big matrices has then lead to the discovery of a twist even
eigenvector with non vanishing $0$--component. This eigenvector is not visible in Siegel
gauge and, as we have seen, it corresponds to the midpoint of the (ghost part of the) string.

%%%%%%%%%%%%%%%%%%%%%%%%%%%%%%%%%%%%%%%%%
\subsection{Diagonalization of  the reduced product}
%%%%%%%%%%%%%%%%%%%%%%%%%%%%%%%%%%%%%%%%%

Once we know the spectrum of the twisted product we can use the
equality of the twisted sliver and the reduced sliver (i.e. sliver in
Siegel gauge) to directly compute the spectrum of the reduced Neumann
coefficients.  Here again we can define ``wedge''-states as

\beq\label{wedge}
|N\rangle=\left(|\dot{0}\rangle\right)_{*_{b_0}}^{N-1}= \N_N
\exp\left(\sum_{n,m=1}^\infty c_n^\dagger\, (CT_N)_{nm}\,
b_m^\dagger\right)|\dot{0}\rangle
\eeq
Which are defined by
\beq
|N+1\rangle = |N\rangle *_{b_0} |\dot 0\rangle
\eeq
The Neumann coefficients $T_N$, are given by\footnote{The expression is formally
identical to the matter case, this is so because the  linear and
non linear properties of the reduced Neumann coefficients are
isomorphic to the matter. }

\beq
T_N=\frac{T+(-T)^{N-1}}{1-(-T)^N}
\eeq
In particular we have
\bea
T_2&=&0\\
T_3&=&Y\\
T_\infty&=&T=T'
\eea
The last equality follows from the fact that the twisted sliver is identical
to the reduced sliver in Siegel gauge.  We recall here that the name ``wedge states''
 is somehow misleading, since this states cannot be interpreted as surface
state in the non twisted CFT, this is so because the star product in
the usual CFT increase the ghost number (as opposed to the twisted
star product). In this sense these states can be properly defined only
algebraically via the reduced product.\\
At $N=3$, we get the eigenvalues of $Y$, $y(\k)$, from the eigenvalues
of $T=T'$ (\ref{tau'})
\beq
y(\k)=\frac{1}{2\cosh\frac{\pi\kappa}{2}-1}
\eeq
Using  (\ref{lprop}) we obtain the spectrum for the
Neumann coefficients $Y_\pm$, which we call $y_\pm(\k)$

\beq
y_\pm(\k)=\frac{\cosh \frac{\pi\kappa}{2} \pm \sinh
\frac{\pi\kappa}{2} -1}{2\cosh \frac{\pi\kappa}{2} -1}
\eeq
%%%%%%%%%%%%%%%%%%%%%%%%%
\section{Conclusions}
%%%%%%%%%%%%%%%%%%%%%%%%%
In this paper we have shown how the zero momentum sector of the matter and the Siegel
gauge part of the ghosts are related by a twist in the CFT of the $bc$ system.\\
The fact that the ghost vertex is related to the matter sector is not a novelty and is
known since \cite{GJ2}, however this correspondence was shown there indirectly via  the 4--strings vertex.\\
We have indeed proven that this correspondence can be understood  by the twist of the $bc$
system and the equality of the twisted sliver and the sliver in Siegel gauge. We have, in
particular, showed that the twisted CFT defines exactly the matter Neumann coefficients up
to a minus sign. The  ``physical'' meaning of this deep relation (which connects two very
different CFT's, one of second order and the other of first order) is probably related to
the fact that the ghosts eliminate the unphysical degrees of freedom of the string, so it
can be expected that there exists a representation of them in which they are  isomorphic to
the non dynamical DF of the matter. This representation is the twist where the central charge
is $-2$, the opposite of  two light--cone matter directions.\\
The slivers of the three sectors are the same (up to signs and symmetrizing factors), even
though they are defined from star products which are manifestly different. This can possibly
be traced back to the fact that these projectors exhibit a singular geometry \cite{MT} and,
in this sense, are insensitive of the underlying CFT. It would be interesting to see if such
equality still holds for more general and less singular projectors like the butterflies and
others studied in \cite{GRSZ2}.\\

\acknowledgments
We thank Loriano Bonora and  Leonardo Rastelli
for useful comments and discussions.

%%%%%%%%%%%%%%%%%%%%%%%%%%%%%%%%%%%%%%%%


\begin{thebibliography}{99}
%%%%%%%%%%%%%%%%%%%%%%%%%%%%%%%%%%%%%%%%

\bibitem{W1} E.Witten, {\it Noncommutative Geometry and String Field
Theory}, Nucl.Phys. {\bf B268} (1986) 253.

\bibitem{Sen} A.Sen {\it Descent Relations among Bosonic D--Branes},
Int.J.Mod.Phys. {\bf A14} (1999) 4061, [hep-th/{9902105}].  {\it
Tachyon Condensation on the Brane Antibrane System} JHEP {\bf 9808}
(1998) 012, [hep-th/{9805170}].  {\it BPS D--branes on
Non--supersymmetric Cycles}, JHEP {\bf 9812} (1998) 021,
[hep-th/{9812031}].
%%CITATION = HEP-TH 9902105;%%
%%CITATION = HEP-TH 9805170;%%
%%CITATION = HEP-TH 9812031;%%

\bibitem{ohmori} K.Ohmori, {\it A review of tachyon condensation in
open string field theories}, [hep-th/0102085].
%%CITATION = HEP-TH 0102085;%%

\bibitem{WT} W.Taylor, {\it Lectures on D-branes, tachyon condensation
and string field theory}, [hep-th/0301094].
%%CITATION = HEP-TH 0301094;%%

\bibitem{leclair1} A.Leclair, M.E.Peskin, C.R.Preitschopf, {\it String
Field Theory on the Conformal Plane. (I) Kinematical Principles},
Nucl.Phys. {\bf B317} (1989) 411.

\bibitem{leclair2} A.Leclair, M.E.Peskin, C.R.Preitschopf, {\it String
Field Theory on the Conformal Plane. (II) Generalized Gluing},
Nucl.Phys. {\bf B317} (1989) 464.

\bibitem{RSZ1} L.Rastelli, A.Sen and B.Zwiebach, {\it String field
theory around the tachyon vacuum}, Adv.\ Theor.\ Math.\ Phys.\  {\bf
5} (2002) 353 [hep-th/{0012251}].
%%CITATION = HEP-TH 0012251;%%


\bibitem{GRSZ1} D.Gaiotto, L.Rastelli, A.Sen and B.Zwiebach, {\it
Ghost Structure and Closed Strings in Vacuum String Field Theory},
[hep-th/{0111129}].
%%CITATION = HEP-TH 0111129;%%

\bibitem{HKw} H.Hata and T.Kawano, {\it Open string states around a
classical solution in vacuum string field theory}, JHEP {\bf 0111}
(2001) 038 [hep-th/{0108150}].
%%CITATION = HEP-TH 0108150;%%

\bibitem{tope} L.~Bonora, C.~Maccaferri, D.~Mamone and M.~Salizzoni,
{\it Topics in string field theory,} [hep-th/{0304270}].
%%CITATION = HEP-TH 0304270;%%

\bibitem{Oku1} K.Okuyama, {\it Siegel Gauge in Vacuum String Field
Theory}, JHEP {\bf 0201} (2002) 043 [hep-th/{0111087}].
%%CITATION = HEP-TH 0111087;%%

\bibitem{Oku2} K.Okuyama, {\it Ghost Kinetic Operator of Vacuum String
Field Theory}, JHEP {\bf 0201} (2002) 027 [hep-th/{0201015}].
%%CITATION = HEP-TH 0201015;%%

\bibitem{RSZ5} L.Rastelli, A.Sen and B.Zwiebach, {\it Star Algebra
Spectroscopy}, JHEP {\bf 0203} (2002) 029 [hep-th/{0111281}].
%%CITATION = HEP-TH 0111281;%%

\bibitem{WT0} Washington Taylor, {\it D--brane effective field theory
from string field theory}, [hep-th/0001201].
%%CITATION = HEP-TH 0001201;%%

\bibitem{KP} V.A.Kostelecky and R.Potting, {\it Analytical
construction of a nonperturbative vacuum for the open bosonic string},
Phys.\ Rev.\ D {\bf 63} (2001) 046007 [hep-th/{0008252}].
%%CITATION = HEP-TH 0008252;%%

\bibitem{belov} D.~M.~Belov, {\it Diagonal representation of open
string star and Moyal product}, [hep-th/{0204164}].
%%CITATION = HEP-TH 0204164;%%

\bibitem{GJ1} D.J.Gross and A.Jevicki, {\it Operator Formulation of
Interacting String Field Theory}, Nucl.Phys. {\bf B283} (1987) 1.

\bibitem{GJ2} D.J.Gross and A.Jevicki, {\it Operator Formulation of
Interacting String Field Theory, 2}, Nucl.Phys. {\bf B287} (1987) 225.

\bibitem{RSZ2} L.Rastelli, A.Sen and B.Zwiebach, {\it Classical
solutions in string field theory around the tachyon vacuum}, Adv.\
Theor.\ Math.\ Phys.\  {\bf 5} (2002) 393 [hep-th/{0102112}].
%%CITATION = HEP-TH 0102112;%%

\bibitem{okuda} T.Okuda, {\it The Equality of Solutions in Vacuum
String Field Theory}, Nucl.\ Phys.\  {\bf B641} (2002) 393
[hep-th/{0201149}].
%%CITATION = HEP-TH 0201149;%%

\bibitem{GRSZ2} D.~Gaiotto, L.~Rastelli, A.~Sen and B.~Zwiebach, {\it
Star Algebra Projectors}, JHEP {\bf 0204} (2002) 060
[hep-th/{0202151}].
%%CITATION = HEP-TH 0202151;%%

\bibitem{MT} G.Moore and W.Taylor {\it The singular geometry of the
sliver}, JHEP {\bf 0201} (2002) 004 [hep-th/{0111069}].
%%CITATION = HEP-TH 0111069;%%

\bibitem{Erler} T.~G.~Erler, {\it ``Moyal formulation of Witten's star
product in the fermionic ghost  sector,''} [hep-th/0205107].
%%CITATION = HEP-TH 0205107;%%

\bibitem{Barso} I.~Bars, I.~Kishimoto and Y.~Matsuo, {\it``Fermionic
Ghosts in Moyal String Field Theory,''} [hep-th/0304005].
%%CITATION = HEP-TH 0304005;%%

\bibitem{Furu} K.~Furuuchi and K.~Okuyama, {\it ``Comma vertex and
string field algebra,''} JHEP {\bf 0109} (2001) 035
[hep-th/0107101].
%%CITATION = HEP-TH 0107101;%%


\bibitem{Belov8} 
D.~M.~Belov and C.~Lovelace, 
{\it ``Star products made easy,''} 
[hep-th/0304158]. 
%%CITATION = HEP-TH 0304158;%% 
 
\bibitem{Kling} 
A.~Kling and S.~Uhlmann, 
{\it ``String field theory vertices for fermions of integral weight,''} 
[hep-th/0306254]. 
%%CITATION = HEP-TH 0306254;%% 

\end{thebibliography}
\end{document}